\documentclass[aps, prl, reprint,graphicx,groupedaddress,amsmath,amssymb]{revtex4-1}
\usepackage{dcolumn}% Align table columns on decimal point
\usepackage{bm}% bold math
\usepackage{natbib}
\usepackage{bibentry}
\usepackage[usenames,dvipsnames] {color}
\usepackage[table]{xcolor}
\usepackage{graphicx}
\usepackage{epstopdf}
\usepackage{amssymb}
\usepackage{amsmath}
\usepackage{dcolumn}
\usepackage{subfigure}
\usepackage[utf8]{inputenc}

\DeclareGraphicsExtensions{.eps,.png,.pdf}

\usepackage{hyperref}
\hypersetup{
colorlinks   = true, %Colours links instead of ugly boxes
urlcolor     = blue, %Colour for external hyperlinks
linkcolor    = blue, %Colour of internal links
citecolor   = blue %Colour of citations
}
\usepackage{setspace}
\usepackage[font=singlespacing,font=footnotesize, justification=RaggedRight]{caption}
\usepackage{enumitem}
\usepackage{fancybox}
\usepackage{comment}
\usepackage{url}
\usepackage{here}
\setlist{nolistsep}
\usepackage{multimedia}
\usepackage{soul}
\setstcolor{red}
\DeclareMathAlphabet      {\mathbf}{OT1}{cmr}{bx}{n}

\newcommand{\printfnsymbol}[1]{%
\textsuperscript{\@fnsymbol{#1}}%
}
\makeatother

\newcommand{\beginsupplement}{%
\setcounter{table}{0}
\renewcommand{\thetable}{S\arabic{table}}%
\setcounter{figure}{0}
\renewcommand{\thefigure}{S\arabic{figure}}%
}

\graphicspath{{./Figures/}{./Figures/}}

\begin{document}

\title{A minimal model of solitons in nematic liquid crystals}

\author{Noe Atzin$^{1}$}
\thanks{equal contribution}
\author{Ali Mozaffari$^{1,2}$}
\thanks{equal contribution}
\author{Xingzhou Tang$^{1}$}
\author{Soumik Das$^{3}$}
\author{Nicholas L. Abbott$^{3}$}
\author{Juan J. de Pablo$^{1,3}$}
\email{depablo@uchicago.edu}

\affiliation{
 $^1$Pritzker School of Molecular Engineering, The University of Chicago, Chicago, Illinois 60637, United States \\
 $^2$OpenEye Scientific, Cadence Molecular Sciences, Santa Fe, New Mexico 87508, USA\\
 $^3$Smith School of Chemical and Biomolecular Engineering, Cornell University, Ithaca, New York 14853, United States\\
 $^4$Center for Molecular Engineering, Argonne National Laboratory, Lemont, Illinois 60439, United States
}

\date{\today}

\begin{abstract}
\begin{singlespace}
Solitons in liquid crystals have generated considerable interest. Several hypotheses of varying complexity have been advanced to explain how they emerge, and a consensus has not emerged yet about the underlying forces responsible for their formation or their structure. In this work, we present a minimal model for soliton structures in achiral nematic liquid crystals, which reveals the key requirements needed to generate traveling solitons in the absence of added charges. These include a surface imperfection or inhomogeneity capable of producing a twist, flexoelectricity, dielectric contrast, and an applied AC electric field that can couple to the director's orientation. Our proposed model is based on a tensorial representation of a confined liquid crystal, and it predicts the formation of "butterfly" structures, quadrupolar in character, in regions of a slit channel where the director is twisted by the surface imperfection. As the applied electric field is increased, solitons (or "bullets") become detached from the wings of the butterfly, which then rapidly propagate throughout the system. The main observations that emerge from the model, including the formation and structure of butterflies, bullets, and stripes, as well as the role of surface imperfections and the strength of the applied field, are consistent with our own experimental findings presented here for nematic LCs confined between two chemically treated parallel plates.   
\end{singlespace}
\end{abstract}

\maketitle

Traveling waves that propagate without distorting their shape or losing energy are referred to as solitons, and they have been studied extensively in isotropic liquids \cite{Zab65,Man04}. Solitons were first described by Russell et al. over 150 years ago (1834), who studied how a wave propagates at constant speed in a narrow channel \cite{Dau06}. The first attempts to understand the topological structures associated with solitons in liquid crystals took place more than 60 years ago \cite{Gen71,Leg72}. In achiral nematic liquid crystals, solitons can be produced by applying an electric field \cite{Ear22,Li19,Aya20,Shen20}. After a collision, solitons preserve their shape, thereby providing opportunities for the design of autonomous soft matter microsystems that take advantage of the energy transduction and active transport \cite{Zha21,Lam92,Hel69}.

Recent experimental work by our group has revealed that surface interactions play a decisive and previously unappreciated role in the generation and dynamics of solitons \cite{Das22}. However, the precise structure of a nematic soliton or the forces that lead to its creation remains unknown. In this work, we propose a minimal model for solitons that reproduces the central features observed in experiments. With that model, we show that surface imperfections serve to nucleate solitons, we find that their topological structure is a function of the applied electric field, and we dissect how the balance of free energy contributions from flexoelectricity, surface anchoring, and elasticity is altered throughout the soliton formation process.

\section{METHODOLOGY}
\subsection{Theoretical Framework}

The theoretical framework adopted here is based on a Landau-de Gennes free energy functional of the tensorial order parameter, $\mathbf{Q}$. For uniaxial systems, it is written in the form of $\mathbf{Q}=S(\mathbf{nn}-\mathbf{I}/3)$, where the unit vector $\mathbf{n}$ is the nematic director field and S, the largest eigenvalue of $\mathbf{Q}$, quantifies the degree of uniaxial alignment \cite{Moz21}. We introduce an alternating current (AC) field that couples to the director via a negative anisotropy of the permittivity of the liquid crystal. The liquid crystal is confined between two walls, and a surface irregularity is created by appending a hemispherical particle to one of the walls. The tensorial order parameter is evolved through the Ginzburg-Landau equation \cite{Arm15a,Arm15b,Ber94,She20}:
\begin{equation}
{\partial \over \partial t} \mathbf{Q}= \Gamma \mathbf{H},
\label{Gin-LdG}
\end{equation}
where $\Gamma$ is a collective rotational diffusion constant that controls the relaxation rate, and $\mathbf{H}$ is the molecular field,
\begin{equation}
\mathbf{H}= - \left({\delta \mathcal{F} \over \delta \mathbf{Q}} - {\mathbf{I} \over 3}  \textrm{Tr} {\delta \mathcal{F}\over \delta \mathbf{Q}}  \right).
\label{MolField}
\end{equation}
The free energy of the system is given by
\begin{equation}
\mathcal{F}= \int_{V} f_{\textrm{bulk}} \textrm{d} V + \int_{S} f_\textrm{surf} \textrm{d}S,
\label{TotEner}
\end{equation}
and the bulk free energy density, $f_\textrm{bulk}$, is given by the sum of an enthalpic contribution ($f_\textrm{LdG}$),  and elasticity contribution ($f_\textrm{elas}$), which accounts for distortions of the nematic from a uniform configuration, a flexoelectricity contribution ($f_\textrm{flex}$), and a dielectric energy contribution ($f_\textrm{diel}$), according to
\begin{equation}
f_\textrm{bulk}=f_\textrm{LdG}+f_\textrm{elas}+f_\textrm{flex}+f_\textrm{diel}.
\label{BulkEne}
\end{equation}
The enthalpic term is given by \cite{Gen95}:
\begin{eqnarray}
f_\textrm{LdG}&=&{A \over 2}\left(1-{U \over 3}\right)\mathrm{Tr} \left(\mathbf{Q} \right)^2- {A U \over 3}\mathrm{Tr}\left(\mathbf{Q}^3 \right) \nonumber \\ 
&+&{A U \over 4}\left(\mathrm{Tr} \left(\mathbf{Q^2} \right) \right)^2,
\label{LdGEne}
\end{eqnarray}
where A and U are phenomenological parameters \cite{Gen69,Cha95}. In the second term, the elastic distortions can be written as \cite{Mor99,Neh72}
\begin{eqnarray}
f_\textrm{elas}&=&{1 \over 2} L_1 \partial_k Q_{ij} \partial_k Q_{ij}+{1 \over 2}L_2\partial_k Q_{jk} \partial_l Q_{jl} \nonumber \\
 &+&{1\over 2}L_3 Q_{ij}\partial_i Q_{kl} \partial_j Q_{kl}+ {1\over 2} L_4 \partial_l Q_{jk} \partial_k Q_{jl}.
\label{ElasEne}
\end{eqnarray}
In this work, we include $L_1$and $L_3$ and assume $L_2=L_4=0$. In the presence of an externally applied electric field, a flexoelectric contribution to the free energy arises from the coupling between the nematic distortion and polarization \cite{Ale93,Blo13}:
\begin{equation}
f_\textrm{flex}= E_i P_i= \zeta_1 \left(\partial_j Q_{ij} \right) E_i+ \zeta_2 Q_{ij}\left(\partial_k Q_{jk} \right) E_i.
\label{FlexoEne}
\end{equation}
The dielectric energy contribution is given by \cite{Lan13,Por11}
\begin{equation}
f_\textrm{diel}=-{1 \over 2} \epsilon_0 \epsilon_{ij} E_i E_j,
\label{DieEne01}
\end{equation}
where $\epsilon_0$ represents the dielectric permittivity of vacuum, and $\epsilon_{ij}$ corresponds to the tensorial dielectric permittivity of the nematic material: $\epsilon_{ij}=\epsilon_\perp \delta_{ij}+\left(\epsilon_\parallel-\epsilon_\perp \right) n_i n_j$; here, $\epsilon_\perp$ and $\epsilon_\parallel$ are the dielectric permittivity perpendicular and parallel to the nematic field, respectively. By introducing an isotropic dielectric permittivity $\bar{\epsilon}$ and permittivity anisotropy $\epsilon_a=\epsilon_\parallel-\epsilon_\perp$, one has $\epsilon_{ij}=\bar{\epsilon} \delta_{ij}+\epsilon_a Q_{ij}$ \cite{Par09,Pal07}. The above expressions can be rewritten in terms of the $\mathbf{Q}$-tensor as:
\begin{equation}
f_\textrm{diel}=-{1 \over 2} \epsilon_0 \epsilon_a E_i Q_{ij} E_j.
\label{DieEne}
\end{equation}
At the confining boundaries, d$S$ has unit normal $\boldsymbol{\nu}$. Anchoring conditions can be imposed by adding a surface term to the free energy \cite{Yok97,Lon15}: 
\begin{equation}
\mathcal{F}_\textrm{surf}= \int_S f_\textrm{surf} \textrm{d}S,
\label{SurfEne}
\end{equation}
where the integration is performed over the confining boundaries, $\partial V$. A hybrid anchoring of the director field is enforced at the surface. Homeotropic anchoring is imposed using a Rapini-Papoular surface free energy density of the form \cite{Rap69,Pap69}
\begin{equation}
f_\textrm{R-P}={1 \over 2} W \left( \mathbf{Q} -\mathbf{Q^0}\right),
\label{HomEne}
\end{equation}
which penalizes deviations from the surface-preferred tensorial order parameter $\mathbf{Q}^0=S_\textrm{eq}\left(\boldsymbol{\nu \nu} - \mathbf{I}/3\right)$. A fourth-order Fournier-Galatola free energy density is adopted to impose degenerate planar anchoring at the walls \cite{Fou05}
\begin{equation}
f_\textrm{F-G}={1 \over 2} W\left(\bar{\mathbf{Q}}-\bar{\mathbf{Q}}_\perp \right)^2 +{1 \over 4} W \left(\bar{\mathbf{Q}}:\bar{\mathbf{Q}} -S_\textrm{eq}^2 \right)^2,
\label{PlaEne}
\end{equation}
where W controls the anchoring strength, $\bar{\mathbf{Q}} = \mathbf{Q} + ({1/3}) S_\textrm{eq} \boldsymbol{\delta}$ and its projection on the surface is given by $\bar{\mathbf{Q}}_\perp= \mathbf{p}\cdot \bar{\mathbf{Q}}\cdot \mathbf{p}$, with $\mathbf{p}=\boldsymbol{\delta -\nu \nu}$.
An effective electric field, which we refer to as \textit{EF}*, can be defined using the ratio between the flexoelectric term and the elastic energy,
\begin{equation}
\textrm{\textit{EF}}^*={\zeta_2 \over L1} E_i.
\label{ElecField}
\end{equation}
This variable can be used to determine the point at which the electric field is sufficiently strong to overcome the elastic energy and alter the director's field.

\begin{figure*}[!htb]
 \centering
\includegraphics[width=0.9\textwidth]{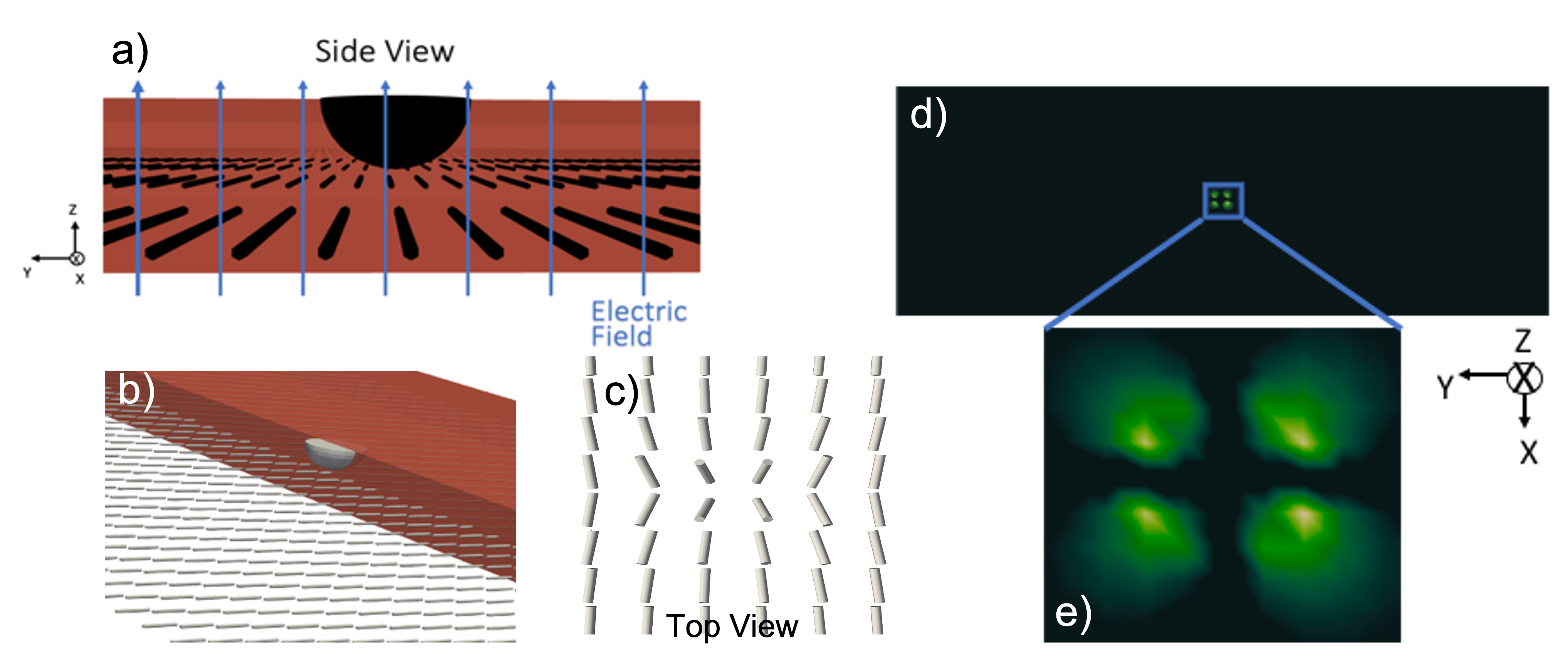}
\caption{Simulated system: (a) the electric field is applied perpendicular to the confining walls, which impart parallel anchoring and (b) a hemispherical particle is placed on the top boundary. (c) The particle generates a small distortion in the director field (d) Cross-polarized light image for the system at equilibrium. (e) Magnified view of the cross-polarized image showing a cross-like pattern that is generated by the homeotropic anchoring on the surface of the particle.}
\label{Fig01}
\end{figure*}

\subsection{Experimental System}
Our experiments were carried out using confining surfaces pre-treated with self-assembled monolayers (SAMs) of alkanethiols on obliquely deposited thin gold films. To prepare the gold substrates, glass slides (Fisher’s Finest glass slides purchased from Fisher Scientific (Pittsburgh, PA)) were placed in an electron beam evaporator (VEC-3000-C manufactured by Tekvac Industries, Brentwood, NY).  The surface normal of each glass slide was tilted by an angle of 45$^{\circ}$ away from the incident direction of the gold vapor. A thin layer of titanium (20 ) was deposited on the glass substrate, followed by a semi-transparent layer of gold (200 \AA). SAMs of 1-hexadecanethiol (C16SH) were prepared by immersing the gold substrates in 2mM ethanolic solutions of 1-hexadecanethiol for 1 hour, followed by rinsing with ethanol and water.  Prior studies have established that SAMs of C16SH cause planar anchoring of CCN-47 \cite{Das22}. Two such treated substrates were placed parallel to each other and separated by 8 $\mu$m spacers to prepare optical cells. Additional details about our experimental procedures can be found \cite{Das22}. We dispersed 3 $\mu$m polyethylene (PE) particles in 4$'$-butyl-4-heptyl-bicyclohexyl-4-carbonitrile (CCN-47), and the mixture was drawn into the optical cell by capillary action at 70$^{\circ}$ (T $>$ TNI), followed by cooling to the nematic phase (T = 45$^{\circ}$C). Polarizing optical microscopy was performed using an Olympus BX41 with crossed polars, and a first-order red plate compensator (Olympus, U-TP530) inserted at 45$^{\circ}$ with the polarizer. By using optical microscopy (transmission mode; crossed polarizers), we determined that the nematic phase of CCN-47 at 45$^{\circ}$C adopts a uniform azimuthal orientation on C16SH SAMs, which is parallel to the direction of deposition of the gold used to form the gold film.

\section{RESULTS AND DISCUSSIONS}

As mentioned above, our model corresponds to a nematic liquid crystal confined within walls that impart strong planar anchoring, which is consistent with our experimental devices. In the absence of this imposed surface inhomogeneity, the director field remains in a homogenous state and the butterfly structure (or any other soliton-like structure) fails to appear. A small hemispherical particle with homeotropic anchoring is placed at the center of the top wall; its radius is one-fifth of the channel thickness (Fig. \ref{Fig01}). This particle creates a small distortion in an otherwise homogenous structure. An oscillatory electric field is applied normal to the confining boundaries (Z-axis). To better represent the experimental conditions and to control the direction of motion of the solitons, white noise is added to the AC electric field. This white noise is superimposed on a constant electric field, and it promotes the formation of a stable soliton structure. The DC offset determines the soliton direction of motion (See Fig \ref{FigSup01} in Sup. Mat.). The nematic is considered to have a negative dielectric anisotropy ($\Delta \epsilon=\epsilon_\parallel-\epsilon_\perp<0$).  The director field is initially aligned along the X-axis, with periodic boundary conditions in the X and Y directions (Fig. \ref{Fig01}b). Since the simulated system is two orders of magnitude smaller than our typical experimental device, the intensity of the electric field and the size of the director structures are different than those used in experiments. However, when properly normalized using the effective electric field (\textit{EF}$^*$), one can see that the experimental and simulated structures exhibit comparable behaviors.
The structure of the nematic director field in the mid-plane of the channel is shown in Figure \ref{Fig01}. Note that, in the absence of an applied electric field, the director in the mid-plane is undisturbed and is aligned along the X-axis.  For a low \textit{EF}$^*$, when the flexoelectric energy is smaller than the elastic energy intensity (\textit{EF}$^*$ $<$1.0), the flexoelectricity contribution is negligible and the orientation of the director remains unchanged. The nematic structure exhibits a cross-like feature, with directors tilted away from the initial plane of orientation (Fig. \ref{Fig01}d). The cross-polarized light micrographs were simulated using a Jones matrix, with the incident polarized wave vector propagated along the Z-axis.

%The structure of the nematic director field in the mid-plane of the channel is shown in Figure \ref{Fig01}. Note that, in the absence of an applied electric field, the mid-plane director is undisturbed and aligned along the X-axis.  For a low\textit{Electric Field}*, when the flexoelectric energy is smaller than the elastic energy intensity (Electric Field* $<$1.0),  the flexoelectricity contribution is negligible and the orientation of the director remains unchanged. The nematic structure exhibits a cross-like feature, with directors tilted away from the initial plane of orientation (Fig. \ref{Fig01}d). The cross-polarized light micrographs were simulated using a Jones matrix, with the incident polarized wave vector propagated along the Z-axis.

\begin{figure*}[!htb]
\centering
\includegraphics[width=0.8\textwidth]{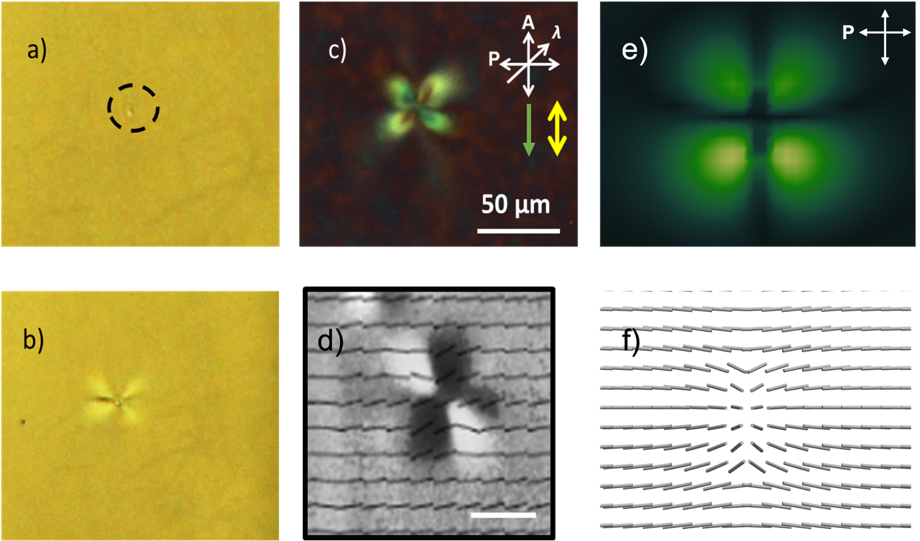}
\caption{2 (a) PE particle (indicated by a black circle) in the LC film confined between two C16SH SAM surfaces. (b-c) The particle nucleates a stationary quadrupolar distortion when an electric field (300 Hz, 40 V) is applied; (b) Bright-field image (c) Imaged under crossed polarizers. The green arrow indicates the direction of Au deposition. The LC alignment is indicated by the yellow double-headed arrow. Scale bars 50 $\mu$m. The results of simulations of the butterfly structure are consistent with the experimental observations. Experimentally, it is observed that the butterfly remains stable in the vicinity of the particle. (d) Polscope image showing the LC director profiles within a stationary butterfly. Scale bar 30 $\mu$m. (e) In simulations, polarized light shows the butterfly in the region of the hemisphere with a 1 $\mu$m size, and (f) the orientation of the director matches the experimental observation.}
\label{Fig02}
\end{figure*}

Upon the application of a slightly higher \textit{EF}$^*$ (\textit{EF}$^*$=1.0), where the flexoelectric energy is comparable to the elastic energy, the flexoelectric energy makes a significant contribution to the nematic liquid crystal around the particle surface; the interplay of flexoelectricity and anchoring leads to the formation of a quadrupolar structure (a butterfly structure). This structure is stable and depends strongly on the inhomogeneity imposed by the hemispherical particle. For the model considered here, butterfly structures are only observed in thin channels where 1.0$<$ \textit{EF}$^*$ $<$1.5 (Fig \ref{Fig02}). The analysis of the director field profile and cross-polarized light simulations show that the director orientation is more distorted and tilted further in the third dimension for the butterfly. As shown in the figure, the simulated polarized light image is consistent with our experimental observations atop a small spherical particle. In all cases, if the electric field is removed, the system returns to the homogenous configuration after 2,000 steps (See Fig \ref{FigSup02}  in Sup. Mat.); we define that time scale as the characteristic relaxation time ($\tau$). 

\begin{figure*}[!htb]
\centering
\includegraphics[width=0.8\textwidth]{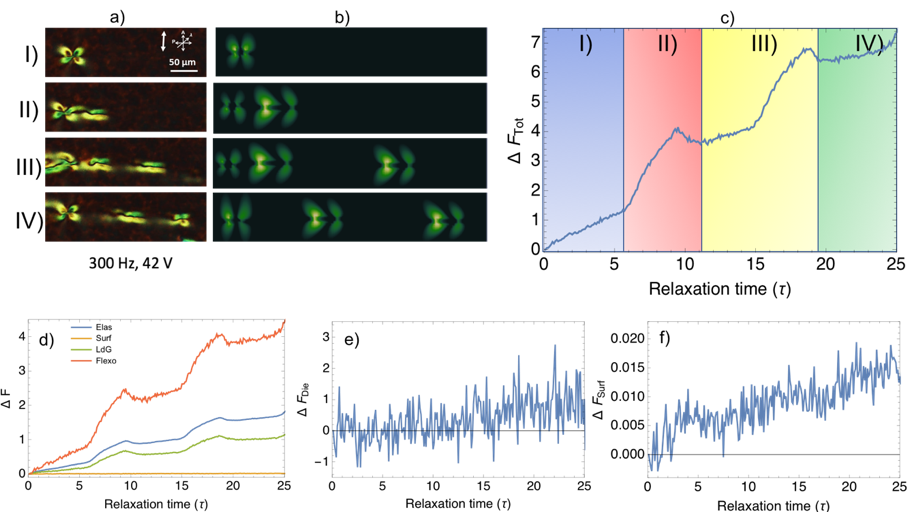}
\caption{(a) Experimental observation of bullet release from a butterfly. (b) Corresponding simulation results, show the creation of a butterfly in the region above the particle, and the repeated ejection of bullets that move at the same speed, in agreement with our experimental observations. (c) Free energy as a function of simulation time; the energy increases until a bullet becomes detached from the butterfly, at which point the energy drops slightly before another bullet is ejected. Each colored region corresponds to the ejection of a new bullet. (d) Landau-de Gennes, elastic and flexoelectric, and surface energy contribution, respectively. (e) The dielectric energy exhibits a small increase throughout the entire process. (f) The surface energy is two orders of magnitude smaller than the other contributions to the free energy; it increases with each bullet that is formed.}
\label{Fig03}
\end{figure*}

Consistent with our experimental observations, for a higher applied \textit{EF}$^*$ (\textit{EF}$^*$ = 1.7 \& 10 $<$ Frequency$*$), the energy of the system starts to increase, and the butterfly adopts an asymmetric structure that tilts toward the positive Y-axis. At some point, the butterfly structure releases a propagating soliton or "bullet", which can be seen in our experiments (Fig. \ref{Fig03}a) and simulations (Fig. \ref{Fig03}b). Figure \ref{Fig03}c shows the free energy of the system (it includes elastic, surface, enthalpic and flexoelectric contributions). When the AC field is applied, the energy of the system gradually increases and, upon the bullet's detachment, it drops and remains almost flat. It then increases again until a new bullet is released. After leaving the butterfly, the bullet moves in a direction perpendicular to the X axis (Fig. \ref{Fig03}b), consistent with our experimental observations (Fig. \ref{Fig03}a). The process continues as new bullets are created and emitted from the butterfly. This topological structure is facilitated by the thin channel because the strong anchoring energy causes the director to return to a homogenous configuration after the bullet advances along the channel; note that for the anchoring strength considered here, the surface contribution is two orders of magnitude smaller than the other contributions to the free energy (Fig. \ref{Fig03} f), but it is sufficient to force the director field to return to the homogenous orientation after the topological structure of the soliton moves away. The other contributions to the free energy (i.e., Landau-de Gennes, elastic, flexoelectric energy) adopt higher values after each bullet is generated. After the bullet becomes detached from the butterfly, the energy drops but not all the way to the butterfly state because the entire system now includes the butterfly and the bullet. With each new bullet the free energy of the system increases (Fig. \ref{Fig03} d). The dielectric energy exhibits a nearly monotonic behavior (except for the random noise) throughout the entire process (Fig. \ref{Fig03} e), and for that reason, it is not included in the total free energy plot shown in Fig \ref{Fig03} c.

\begin{figure*}[!htb]
\centering
\includegraphics[width=0.8\textwidth]{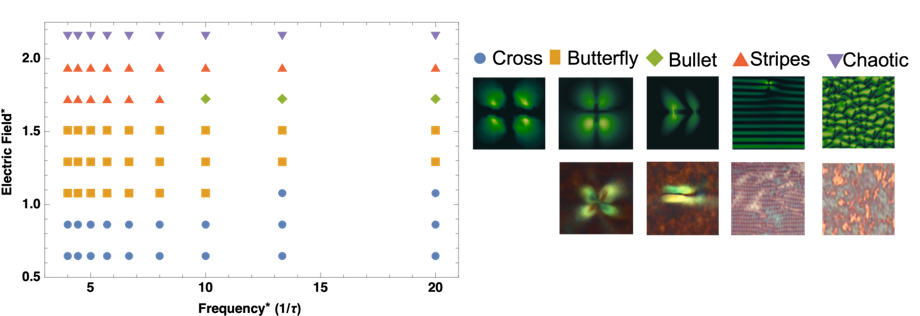}
\caption{(a) The phase diagram serves to delineate the regions where different structures are observed as a function of the frequency and intensity of the electric field. In (b) we show the director's corresponding theoretical and experimental images under cross polarizers; starting from low intensity, we observe a cross pattern above the particle. After increasing the intensity, a butterfly is observed and bullets are generated in a narrow region of frequency. Stripes are observed for a high-intensity field over a wide range of frequencies. The system enters a chaotic regime when the electric field is even stronger. The bottom panel shows the corresponding experimental figures.}
\label{Fig04}
\end{figure*}

\begin{figure*}[!htb]
 \centering
 \includegraphics[width=0.5\textwidth]{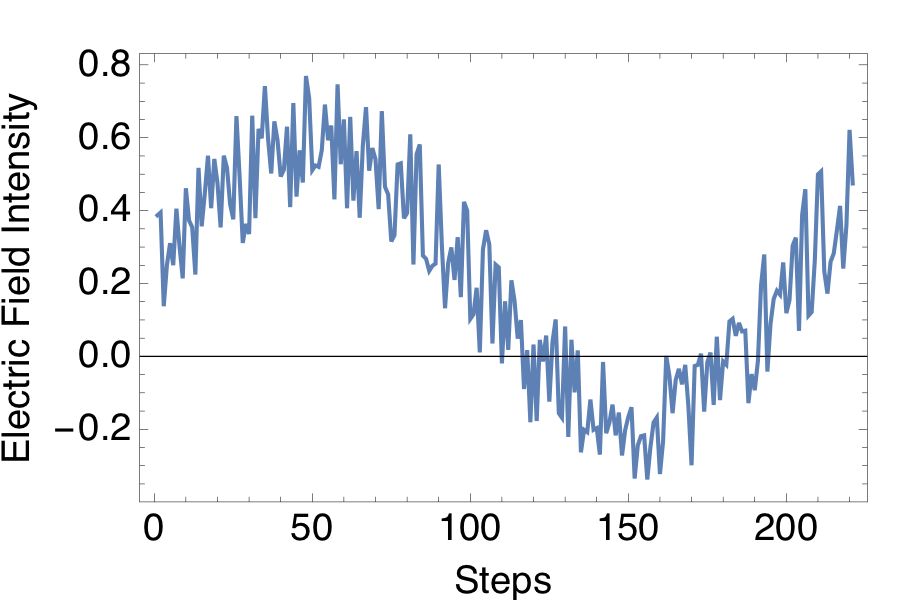}
 \caption{\textbf{AC Electric Field implementation.} The simulations are implemented using an AC electric field decorated with white noise (0,0.1). We shift the electric field in order to break the symmetry contribution to the flexoelectric term and induce motion in the bullets along the positive y-axis. If we add a negative white noise (0,0.1) the direction of motion is towards the negative y-axis.}
 \label{FigSup01}
\end{figure*}

\begin{figure*}[!htb]
 \centering
 \includegraphics[width=0.7\textwidth]{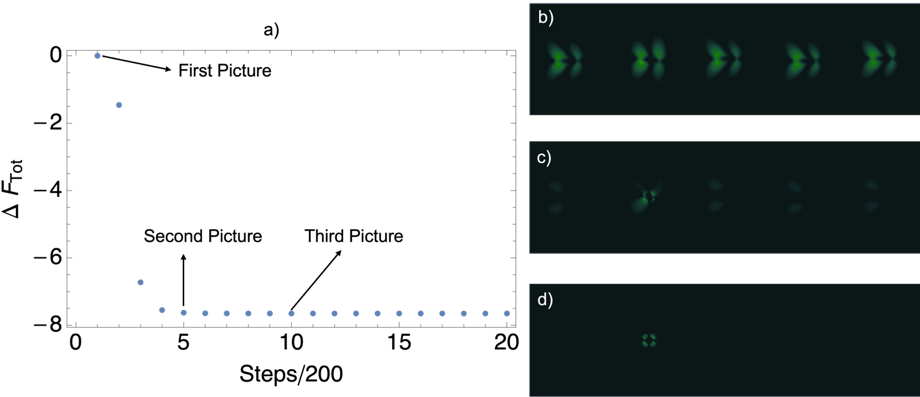}
 \caption{\textbf{Relaxation Time.} The relaxation time is taken as the number of steps that the system needs to return to the homogenous configuration. The simulation starts with the bullets moving along the channel; the electric field is turned off in the first step. In (a) the total free energy shows a decay from the bullet state to the homogeneous state. (b) Starting with the bullets, the electric field is turned off and, after 1,000 steps, the system shows small distortions where the bullets' structures used to be. (d) After 2,000 steps, the system adopts a homogeneous state, and the energy shows a monotonic and stable behavior. The same relaxation occurs if we start with stripes or from a chaotic behavior.}
 \label{FigSup02}
\end{figure*}

\begin{figure*}[!htb]
 \centering
 \includegraphics[width=0.7\textwidth]{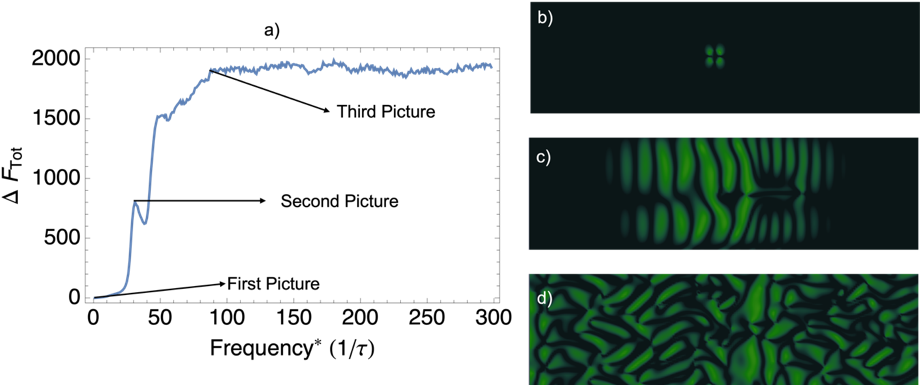}
 \caption{\textbf{Chaotic state energy.} The energy in the chaotic state is three orders of magnitude higher than that in the bullet state, as shown in (a). In (b), we show the start of the simulation. The simulation starts with a homogeneous state and no electric field. (c) Before the chaotic state is simulated, the system shows a mixed state, between a bullet and stripes, where the energy is two orders of magnitude higher than in the bullet state. Finally, in (d), the system enters a chaotic state, and the energy exhibits a monotonic behavior.}
 \label{FigSup03}
\end{figure*}

It is of interest to determine the response of the system across a wide spectrum of applied electric fields. Figure \ref{Fig04} shows a phase diagram, which outlines the five structures that our model predicts as a function of the Frequency$^*$ and intensity of the Electric Field*. Starting from a low intensity (\textit{EF}$^*$ $<$ 1.0), we find that the anchoring of the particle determines the orientation of the director, leading to a cross-pattern under cross-polarizers. Upon increasing the intensity (1.0 $<$ \textit{EF}$^*$ $<$ 1.5), we observe that a "butterfly" is formed, where the flexoelectric contribution begins to change the orientation of the director. Above a threshold, \textit{EF}$^*$ strength (\textit{EF}$^*$ = 1.7 \& 10 $<$  Frequency*), a soliton ("bullet") is formed, which travels in the direction perpendicular to the initial director orientation. When the intensity of the \textit{EF}$^*$ is increased (1.8 $<$ \textit{EF}$^*$ $<$ 2.0) even more, the flexoelectric contribution distorts the entire system homogeneously, and lines ("Stripes") are formed in the direction parallel to the orientation of the director; the stripes move perpendicularly to the director. For an even stronger \textit{EF}$^*$ (2.0 $<$ \textit{EF}$^*$), the system enters a chaotic regime and the free energy is two orders of magnitude higher than the energy of the bullet (See Fig \ref{FigSup03} in Sup. Mat.). Our experiments confirm the general predictions of the theory; a butterfly is formed above the PE microparticle, as can be seen in Figs. \ref{Fig02} and \ref{Fig04}. Above a threshold strength of the \textit{EF}$^*$ bullets are released from the butterfly, as shown in Figs. \ref{Fig03} and \ref{Fig04}. For higher strengths (twice the critical \textit{EF}$^*$), we observe stripes and, eventually, a chaotic regime (see Fig.\ref{Fig04}).

\section{CONCLUSION}
The minimal model adopted here includes contributions from surface energy, elasticity, and flexoelectricity. Our experimental system consists of nematic materials without added salts. While we did not add any ionic species to our experiments, we acknowledge that it is difficult to remove trace amounts of ions in LC films. The results of our simulations, however, indicate that solitonic structures can be created in the absence of ionic effects. A surface inhomogeneity (a hemispherical particle) is introduced to generate an initial distortion of the director field, which serves to "seed" the formation of a butterfly. With that model, we show that upon application of an AC field it is indeed possible to generate a butterfly structure that, above a threshold intensity value, can eject solitons or bullets that travel with uniform velocity throughout the system. The butterfly and the bullets have a quadrupolar structure. The model is then used to generate a non-equilibrium state diagram. The general phenomenology predicted in simulations is consistent with our own experimental observations, which indicate that a butterfly is formed on top of a microparticle that exhibits homeotropic anchoring. For sufficiently strong fields, bullets are ejected from the wings of the butterfly, and they travel at uniform speed throughout the system. At higher field strengths, our simulations and experiments reveal the formation of stripes. Eventually, experiments and simulations show a chaotic regime for high field strengths.
  	Our results indicate that flexoelectricity plays a key role in the formation of solitons. At rest, the butterfly is stabilized by the balance between flexoelectricity, Landau-de Gennes enthalpic energy, anchoring, and elasticity. Above a threshold field intensity, that balance is altered - the flexoelectric contribution rises and the total energy increases to a maximum when the incipient bullet touches the surface of the hemispherical particle. At that moment, the bullet becomes detached from the particle, and the total energy decreases again until a new bullet is generated.  
The bullets are a function of the channel size, frequency, and intensity of the electric field. For our model, they are only observed over a narrow region of field intensity and frequency, and they are susceptible to the nature of the surface imperfection (size, strength anchoring, and anchoring type of the hemispherical particle). Indeed, bullet structures are primarily observed in thin channels. Thick channels generally result in chaotic behavior. 
Solitons provide opportunities for fast transport in liquid crystals, which could lead to a new generation of microfluidic devices for separations, sensing, or logic operations. The work presented here serves to define their structure, and the conditions necessary to generate them.

\bibliography{Soliton.bib}

\ \\

\ \\

\ \\

\section*{Acknowledgments}
This work is supported by the Department of Energy, Basic Energy Sciences, Division of Materials Science and Engineering, Biomaterials Program, under grant no. DESC0019762

\clearpage

\begin{center}
 \section*{Supplementary Information}\label{SI}
\end{center}
\beginsupplement

\subsection{Numerical details}
We solve the Ginzburg-Landau equation using a finite-difference method \cite{Arm15b,Oss06,Kum18,Guo13}. The coherence length ($\xi$) is 10 nm, which we take as the unit of length, and is typical of a nematic liquid crystal \cite{Atz18}. The other parameters are chosen to be $A = 0.1$, $L_1= 0.0116$, $L_3=-0.001$, $\zeta_1=0.01$, $\zeta_2=0.05$, $\epsilon_a=-0.01$, $\Gamma = 0.6$ and $U = 3.5$, which are also representative of the liquid crystal considered in our experiments. The simulation is performed on a square lattice with periodic boundary conditions along the x and y axes, and with planar anchoring in the top and bottom walls. The dimensions are [Nx, Ny, Nz ]= [60, 400,10]; a hemi-spherical particle is placed in the center with a radius of 3.

\subsection{AC Electric Field}

\subsection{Relaxation Time}

\subsection{Chaotic state energy}

\end{document}